\documentclass[fleqn,usenatbib]{mnras}

\usepackage{newtxtext,newtxmath}

\usepackage[T1]{fontenc}
\usepackage{ae,aecompl}

\usepackage{graphicx}	
\usepackage{amsmath}	

\title[An inflated, weakly irradiated brown dwarf]{NLTT5306B: an inflated, weakly irradiated brown dwarf}

\author[S. L. Casewell et al]{S. L. Casewell$^{1,\dagger}$ \thanks{E-mail: slc25@le.ac.uk},  J. Debes$^{2}$, I. P. Braker$^{1}$, M. C. Cushing$^{3}$, G. Mace$^{4}$, \newauthor M.~S. Marley$^{5}$ and J.Davy Kirkpatrick$^{6}$\\
$^{1}$School of Physics and Astronomy,  University of Leicester,
University Road, Leicester LE1 7RH, UK \\
$^{2}$pace Telescope Science Institute, Baltimore, MD 21218, USA\\
$^{3}$The University of Toledo, 2801 West Bancroft Street, Mailstop 111, Toledo, OH 43606, USA\\
$^{4}$McDonald Observatory and the Department of Astronomy, The University of Texas at Austin, Austin, TX 78712, USA\\
$^{5}$NASA Ames Research Center, California, USA\\
$^{6}$IPAC, Mail Code 100-22, Caltech, 1200 E. California Blvd., Pasadena, CA 91125, USA\\
$^{\dagger}$ STFC Ernest Rutherford Fellow\\
}

\date{Accepted XXX. Received YYY; in original form ZZZ}

\pubyear{2020}

\begin{document}
\label{firstpage}
\pagerange{\pageref{firstpage}--\pageref{lastpage}}
\maketitle

\begin{abstract}
We present $Spitzer$ observations at 3.6 and 4.5 microns and a near-infrared IRTF SpeX spectrum of the irradiated brown dwarf NLTT5306B. We determine that the brown dwarf has a spectral type of L5 and is likely inflated, despite the low effective temperature of the white dwarf primary star. We calculate brightness temperatures in the $Spitzer$ wavebands for both the model radius, and Roche Lobe radius of the brown dwarf, and conclude that there is very little day-night side temperature difference. We discuss various mechanisms by which NLTT5306B may be inflated, and determine that while low mass brown dwarfs (M$<$35 M$_{\rm Jup}$) are easily inflated by irradiation from their host star, very few higher mass brown dwarfs are inflated. The higher mass brown dwarfs that are inflated may be inflated by magnetic interactions or may have thicker clouds.

\end{abstract}

\begin{keywords}
white dwarfs, brown dwarfs, binaries: close
\end{keywords}



\section{Introduction}
Due to the phenomenon known as the brown dwarf desert \citep{metchev, grether} there are very few known detached post-common envelope binaries comprised of a white dwarf and a brown dwarf. Despite predictions to the contrary, there are also very few interacting systems (cataclysmic variables and polars) containing a brown dwarf (e.g. \citealt{burleigh06, hernandez}).  

One system that straddles the boundary between these non-interacting and interacting systems is NLTT5306AB.  It was first identified as a candidate post-common envelope system by \citet{steele11} and \citet{girven11}, and confirmed by \citet{steele13} who determined the spectral type of the brown dwarf to be between L4 and L7, and a minimum mass to be 56$\pm$3 M$_{\rm Jup}$.

NLTT5306AB is one of five systems known with a period of around two hours (P=101.88 minutes \citealt{steele13}), similar to the 116 minutes of WD0137-349AB \citep{longstaff17} and 121 minutes of SDSSJ141126.20+200911.1 \citep{beuermann13}. Indeed, until the recent discovery of three very short period white dwarf-brown dwarf binaries from the $Kepler$ K2 mission \citep{parsons17, casewell18} with periods of $<$72~mins, NLTT5306AB was the shortest period white dwarf-brown dwarf binary known. 

 As with many of these post-common envelope systems the white dwarf has a low mass (0.44$\pm$0.04 M$_{\odot}$), indicating that during post-main sequence evolution, the giant prematurely ejected its envelope once it engulfed the brown dwarf (e.g. \citealt{marsh95}).   What is particularly interesting about this system is that although the white dwarf is very cool (T$_{\rm eff}$=7756 $\pm$ 35 K), H$\alpha$ emission is seen \citep{steele13, longstaff18}. This emission appears to be emanating from the surface of the white dwarf, and not the brown dwarf. In \citet{longstaff18} we examined optical spectra from NLTT5306AB in more detail and determined that the H$\alpha$ emission line and a previously unseen Na absorption feature were due to accretion onto the white dwarf from the brown dwarf atmosphere.  We did not detect any radio or X-ray emission, suggesting that the interaction that is occurring is likely to be very weak, and while the mass and radius (from evolutionary models) of the brown dwarf suggest it is not Roche lobe filling, the accretion is likely due to a wind, possibly magnetically funnelled onto the white dwarf. Interestingly this accretion rate is similar to the rate at which highly irradiated exoplanets are losing their atmospheres (e.g. \citealt{spake}).
 
 H$\alpha$ emission has been detected in other irradiated brown dwarf systems, both WD0137-349A (T$_{\rm eff}=16500\pm$500 K; \citealt{maxted06}) and EPIC212235321A (T$_{\rm eff}=24490\pm$194 K; \citealt{casewell18}) show emission from H$\alpha$ and other species (e.g. Na, K, Ca) in their spectra. However, unlike in NLTT5306AB, these emission lines move in antiphase to the white dwarf absorption features and are attributed to a chromosphere on the brown dwarf caused by the irradiation from the white dwarf. SDSSJ141126.20+200911.1 has a slightly longer period than WD0137-349AB by $\sim$5 minutes, and the white dwarf is 3500~K cooler (T$_{\rm eff}=13000\pm300$ K: \citealt{littlefair14}) than  WD0137-349AB, but no emission from H$\alpha$ or any other species is seen, suggesting that the white dwarf needs to be hotter than 13000~K to induce a chromosphere. The lower effective temperature of NLTT5306A suggests that there should be no chromospheric emission seen from NLTT5306B which is consistent with the literature.  

We present here a near-IR spectrum of the brown dwarf NLTT5306B and mid-IR lightcurves of the system.  We use the lightcurves to determine the day- and night-side brightness temperatures for this weakly irradiated brown dwarf and discuss the radius of the brown dwarf in comparison to other irradiated brown dwarfs. 

\section{Observations and data reduction}

\subsection{NIR spectroscopy}
On the night of 2012 January 30 we obtained 20 spectra with 120~s exposures using the SpeX spectrograph \citep{rayner03} on the NASA InfraRed Telescope Facility (IRTF) with the 0.5" slit and the prism mode, covering 0.6 to 2.5 microns at a resolution of $\sim$120.

The data were reduced using the \textsc{spextool} reduction package \citep{cushing04} which also performs the telluric correction using an A0 dwarf telluric standard star \citep{vacca03}. The reduced spectrum can be seen in Figure \ref{fig:spex} with a model white dwarf spectrum for reference.

\subsection{\textit{Spitzer} Photometry}
We obtained Spitzer IRAC \citep{fazio04} warm
photometry from Cycle 9 (Programme ID:40325, PI:
Casewell) to determine the nature of any mid-IR excess and
reflection effect present, similar to our observing strategy for WD0137-349AB \citep{casewell15}. Each IRAC channel ([3.6], [4.5] microns)
was observed for one 101 minute orbit. The data were observed
using 30~s integrations, the full array and no dithering,
as time series photometry was required. We performed
a peak up at the start of the observation to ensure the target
was placed on the sweet spot of the detector.
For both wavebands, aperture photometry was performed
on each individual image using the apex software
and an aperture of three pixels with a background aperture of
12-20 pixels. Pixel phase, array location dependence, and
aperture corrections were applied to the data and the IRAC
zero magnitude flux densities as found on the $Spitzer$ website
were then used to convert the flux into magnitudes on
the Vega magnitude scale. The photometric errors were estimated
using the Poisson noise given by the apex software which was then combined with the errors on the zeropoints. The 3 per cent absolute calibration error was added in quadrature
to these photometric errors.

\section{Results}

\subsection{NIR Spectroscopy}

\begin{figure}
	\includegraphics[trim=2cm 2cm 0cm 2cm,clip,scale=0.35]{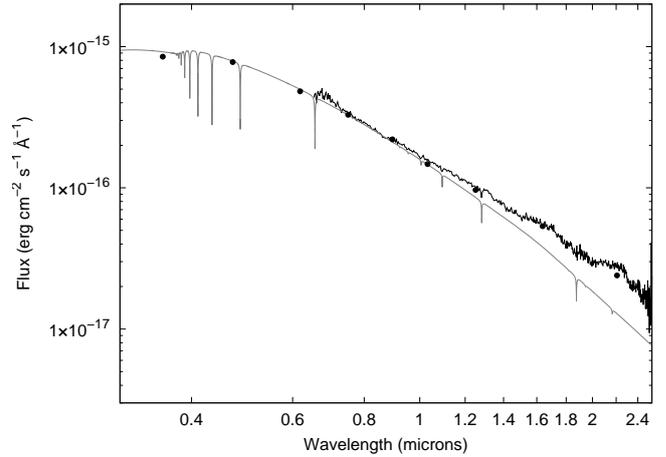}
    \caption{Reduced IRTF spectrum of NLTT5306AB (black) and the model of the white dwarf NLTT5306A (grey: T$_{\rm eff}$=7756 K, log g=7.68) both normalised to the $i$ band photometry before the white dwarf model was subtracted from the data. The SDSS and UKIDSS photometry is also shown and the near-IR excess due to the L dwarf companion can clearly be seen.}
    \label{fig:spex}
\end{figure}

We normalised the combined, calibrated, SpeX spectrum to the white dwarf's $r'$ magnitude from the SDSS catalogue, and a pure hydrogen model of the white dwarf ($T_{\rm eff}$=7756 K, log $g$=7.68) from \citet{koester10} to the same magnitude (Figure \ref{fig:spex}.)
We were then able to subtract the white dwarf model spectrum, convolved to the resolution of SpeX, from the SpeX spectrum to leave a spectrum of the brown dwarf. \citet{steele13} measured the peak to peak amplitude of the lightcurve of NLTT5306 to be only 0.8 per cent in the $i$ band, and we would expect this variation to be smaller in the $r$ band, meaning the dominant errors on this subtraction will be the uncertainty in the $T_{\rm  eff}$ and log $g$, not any reflection effect.

We then normalised the spectrum to 1 in the $J$ band as was done in \citet{burgasser10} to enable comparisons with brown dwarf template spectra.

From comparison to standard L dwarf spectra (Figure \ref{fig:spex2}), we determined the most likely spectral type for the secondary was L5.

\subsection{Indications of low gravity}
We used the low gravity indices of KI$_{J}$, FeH$_{z}$ and $H-cont$ from \citet{allers13} to investigate the spectrum of the brown dwarf.  The indices are measured using the flux at a specific line, and at two continuum wavelengths, all of fixed width. These indices are then converted using the spectral type of the object to give a gravity score of 0 if the indices are consistent with those for objects on the field dwarf sequence, 1, if the indices are 1$\sigma$ away from the field sequence, indicating intermediate gravity ($<$200 ~Myr), and 2 if the indices strongly indicate low gravity (e.g. $\sim$ 10 Myr). The median of gravity scores are then used to determine the final score as it is not a requirement for all the scores to be the same. The line and continuum wavelengths from Table 4 of \citet{allers13} are shown in Table \ref{table:indices} with their calculated indices and the corresponding low gravity score for an L5 dwarf.  

\begin{table*}		
\begin{center}
\begin{tabular}{lccccc}
\hline
Parameter & $\lambda_{line}$($\mu$ m)& $\lambda_{cont1}$ ($\mu$ m)& $\lambda_{cont2}$ ($\mu$ m)&Index & score\\
\hline
KI$_{J}$ &1.244 & 1.220& 1.270&1.043$\pm$0.040&2\\
FeH$_{z}$&0.998 &0.980 &1.022&1.291$\pm$0.0911&0\\
$H-cont$&1.560& 1.470& 1.670&0.934$\pm$0.043 & 1\\
\hline
\end{tabular}
\end{center}
\caption{The gravity sensitive indices for NLTT5306 using the method in  \citet{allers13}.}
\label{table:indices}
\end{table*}

\begin{figure}
	\includegraphics[trim=2cm 2cm 0cm 2cm,clip,scale=0.35]{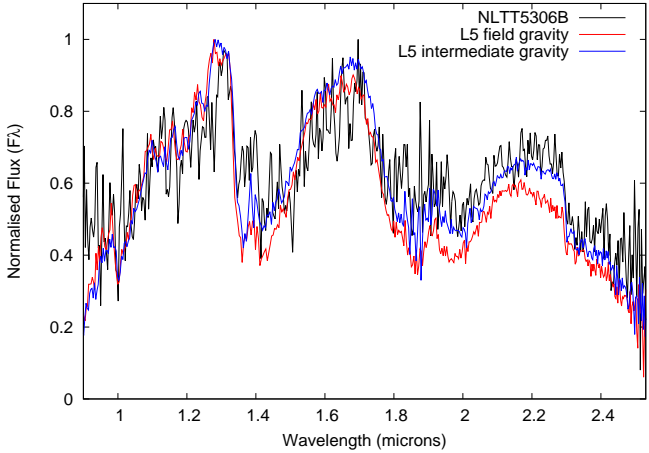}
    \caption{Spectrum of NLTT5306B (black) and two L5 dwarfs. SDSSJ083506.16+195304.4 \citep{chiu06} with field gravity (red) and 2MASSJ0028208+224905 \citep{burgasser10} with intermediate gravity (blue).}
    \label{fig:spex2}
\end{figure}

The gravity scores present in Table \ref{table:indices} have been compared to those for an L5 dwarf as given in \citet{allers13}. Interestingly these values indicate that NLTT5306 has intermediate gravity, similar to that of a L5 dwarf with an age of 50-200 Myr. 

Figure \ref{fig:spex2} shows the NLT5306B spectrum, with two comparison objects. Both are L5 dwarfs, but one has field gravity, and the other intermediate gravity. It can be seen that NLTT5306B has more absorption in the $H$ band than both the field dwarf, and the lower gravity object. There is also more absorption within the blue side of the $J$ band peak at 1 micron within NLTT5306B than in either of the two standard star spectra.   The $K$ band flux is however brighter than that of the field dwarf, and slightly brighter than the intermediate gravity object.  We see brightening in the $K$ band lightcurves of WD0137-349B and SDSS1411B \citep{casewell18, casewell15} compared to models of irradiated brown dwarfs, and so this discrepancy could be due to irradiation induced emission. However, the resolution of the SpeX spectrum is too low to search for H$_{3}^{+}$ emission that could be causing this brightening. 

As the cooling age of NLTT5306A is 710$\pm$50 Myr, and the likely system age $>$ 5 Gyr \citep{steele13}, it is clear the brown dwarf in this system cannot be 50-200 Myr old, the usual age estimate for objects of intermediate gravity which are still shrinking as they cool. However, the signs of intermediate gravity suggest that the brown dwarf may have a smaller log g, and hence larger radius than would normally be suggested for an object of this age.  Using the 5~Gyr Sonora Bobcat models of \citet{marley} a 50 M$_{\rm Jup}$ brown dwarf has a log g of 5.25 and a radius of 0.0901 R$_{\odot}$. This brown dwarf has $T_{\rm eff}$=1500~K, not dissimilar to our measured spectral type of L5. However, such a brown dwarf would not show signs of intermediate gravity in its spectrum. For reference, the same models predict an intermediate gravity brown dwarf (of age 200 Myr) and the same mass would have log g = 5.00 and a radius of 0.11~R$_{\odot}$, similar to the Roche lobe of the brown dwarf. These models also predict such a brown dwarf would be much hotter than an L5 dwarf at $\sim$2000~K. 
We conclude that this brown dwarf is probably inflated. The true radius of the brown dwarf  must lie between the estimate of the Roche lobe (as the lack of strong accretion suggests the system is not Roche lobe filling) and the estimate of the radius from the Sonora Bobcat models at 5~Gyr.

\subsection{Mid-IR lightcurves}
\begin{figure}
	\includegraphics[trim=5cm 2cm 0cm 2cm,clip,scale=0.2]{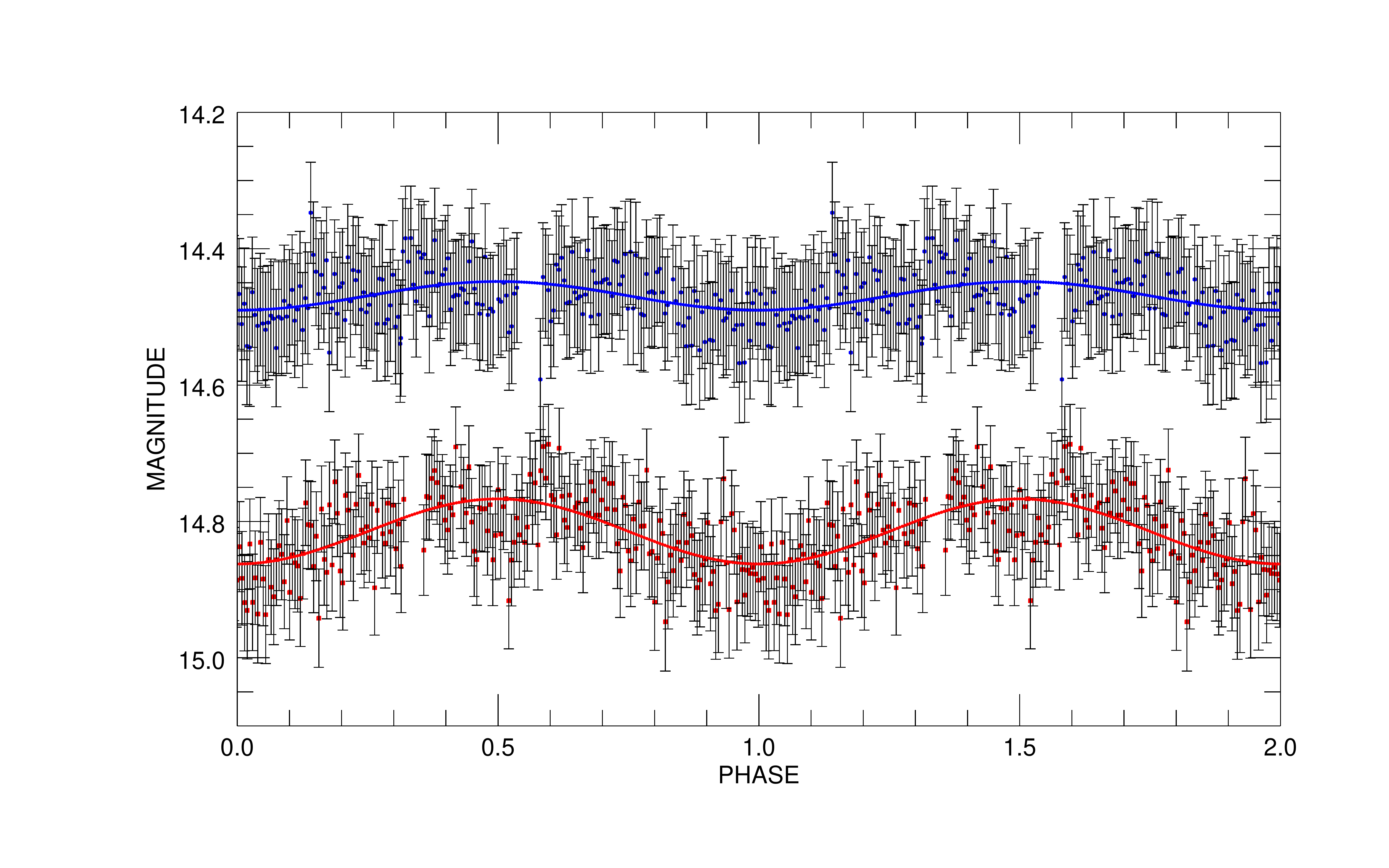}
   \caption{$Spitzer$ IRAC lightcurves in the [3.6] micron (blue) and [4.5] micron (red) bands. The sine curves fitted to the data have been overplotted, and the [3.6] micron data has been offset by -0.25 mags for display purposes. The data have been duplicated over two orbits for display purposes.}
    \label{fig:spitzer}
\end{figure}

In total, after data reduction we have 190 data points in each of the [3.6] and [4.5] micron bands spanning a whole 101 min orbit of the binary (Figure \ref{fig:spitzer}). The gap in the data at phase $\sim$0.6 in the [3.6] micron band is real as we do not have coverage over the entire orbit in this band. The data were phased on the ephemeris given in \citet{longstaff18}, and semi-amplitudes of the reflection effect were determined to be 0.021$\pm$0.009 in the [3.6] micron band and 0.047$\pm$0.007 in the [4.5] micron band. The semi-amplitude of the variability in the $i'$ band was determined to be $\sim$0.5 per cent by \citet{steele13} indicating the mid-IR variability is of larger amplitude. These results, although of lower amplitude, are consistent with those we presented for WD0137-459AB in \citet{casewell15}, where the reflection effect increased with increasing wavelength.

\subsection{Brightness temperatures}
We calculated the brightness temperatures for the day and night sides of NLTT5306 by using the method described in \citet{casewell15}. We used the same white dwarf model as for the SpeX data, and convolved it with the IRAC filter profiles to calculate the white dwarf magnitudes in these wavebands. As an additional check we confirmed they were consistent with the model white dwarf magnitudes in \citet{holberg} when combined with the Gaia distance to the source of 76.96$^{+0.55}_{-0.53}$ pc \citep{gaia}. We then converted all magnitudes to flux and subtracted the white dwarf flux from that of the combined system at the maximum and minimum, to leave the flux from the brown dwarf alone. These values (in magnitudes) are given in Table \ref{table:mags}.

\begin{table*}
\caption{Magnitudes for the combined system, white dwarf and day and night side of the brown dwarf.}

\begin{center}
\begin{tabular}{c c c c c c }
\hline
Waveband &\multicolumn{2}{c}{Magnitude (WD+BD) }& Magnitude (WD)& \multicolumn{2}{c}{Magnitude (BD)}\\
& Dayside& Nightside&&Dayside & Nightside \\ 
\hline
3.6&14.902$\pm$0.084& 14.948$\pm$0.080&16.029 &15.230$^{+0.121}_{-0.110}$&15.293$^{+0.117}_{-0.106}$\\
\\
4.5&14.767$\pm$0.066&14.862$\pm$0.062&16.035 &15.064$^{+0.091}_{-0.084}$&15.192$^{+0.087}_{-0.081}$\\
\hline
\end{tabular}
\end{center}
\label{table:mags}
\end{table*}

Using the method described in \citet{casewell15}, we then converted the brown dwarf fluxes into radiance using the Gaia distance to the system and two assumed radii,  R=0.095 R$_{\odot}$ from \citet{steele13} (assuming it is a field age non-inflated object), and R=0.12 R$_{\odot}$ (assuming the brown dwarf has the radius equal to its  Roche lobe  \citealt{longstaff18}). In reality, the radius is likely to be between these values. The results are in Table \ref{table:brightness}.

\begin{table}
\caption{Brightness temperatures for the day and night sides of the brown dwarf at the model radius of NLTT5306 from \citet{steele13} and assuming the Roche lobe radius. }

\begin{center}
\begin{tabular}{c c c c c }
\hline
Waveband &\multicolumn{4}{c}{Brightness Temperature (K)}\\
 &\multicolumn{2}{c}{R=0.095 R$_{\odot}$}&\multicolumn{2}{c}{ R=0.12 R$_{\odot}$}\\
Microns& Dayside& Nightside&Dayside & Nightside \\ 

\hline
3.6&1756$^{+72}_{-74}$&1717$^{+67}_{-70}$&1478$\pm$54&1449$^{+49}_{-52}$\\
\\
4.5&1651$^{+58}_{-60}$&1568$^{+51}_{-52}$&1361$\pm$42&1302$\pm$38\\
\hline
\end{tabular}
\end{center}
\label{table:brightness}
\end{table}
The errors on both the  [3.6] and [4.5] micron measurements are too high to determine a day-nightside temperature difference with any confidence, although such a difference appears to reside between 60 and 80 K in the [4.5] micron bounds as the errors only just overlap at the upper and lower limits.  Our IRAC photometry shows that while there is a reflection effect seen in this system, it is small, and there is an almost negligible day-night side temperature difference for the brown dwarf.  This is perhaps not surprising as the white dwarf is very cool (7756 K) - in fact the coolest white dwarf known to have a close brown dwarf companion. It is clear however, that if NLTT5306B is significantly inflated, that the brightness temperatures we calculate are considerably lower (by about 300 K), than if it has the smaller radius of a field dwarf.

\section{Discussion}

Using the Gaia distance to the system, we can compare the magnitudes for the brown dwarf alone (Table \ref{table:mags}) to those given in \citet{dupuy}, and determine that they are consistent with L3-L5 spectral type in the both the [3.6] and [4.5] micron bands. This spectral type estimate is earlier than the spectral types of L4-L7 determined in \citet{steele13} from spectral indices. The $K$ band photometry alone however suggests a spectral type more consistent with L4-L5 for the brown dwarf. This is also consistent with the spectral type determined from the SpeX spectrum presented in this work.  There is no considerable mid-IR excess seen in the IRAC bands indicative of brightening due to UV irradiation as has been suggested for WD0137-349B \citep{casewell15}, but due to the much lower temperature of NLTT5306A ($\sim$7000~K compared to 16500~K), this is perhaps not surprising. What is then surprising is the signs of low gravity from the SpeX spectrum indicating NLTT5306B appears to be inflated in some way. In \citet{longstaff18} we suggested that this inflation may be allowing a wind to be  magnetically funnelled onto the white dwarf, however the mechanism for the inflation is still unknown.

We also consider here the possibility that the brown dwarf may not be completely inflated, but may instead be oblate due to either the fast rotation rate of the tidally locked brown dwarf (P=101~mins), or due to the gravitational pull of the white dwarf.
None of the eclipsing systems with a white dwarf primary show any sign of oblateness or deformation in the lightcurves, but we are viewing NLTT5306 at a lower inclination which may make a difference.

The tidal distortion due to the white dwarf was calculated using \textsc{roche}, which also includes the distortion due to the tidally locked rotation, and using the mass ratio, separation and assumed radius of the brown dwarf (0.095 M$_{\odot}$), this total distortion is 2.5 per cent.
Using the moment of inertia for a 56 M$_{\rm Jup}$ brown dwarf at 6~Gyr from the Sonora models \citep{sonora_web}, we can calculate how much of this distortion is due to the rotation rate of NLTT5306B. We used the equations in \citealt{marley} and \citealt{barnes} to determine $f$, the fractional change between the equatorial and polar radius due to the rotation rate of NLTT5306B. This change is 2.5 per cent, or $\sim$1600~km meaning the main source of distortion is due to the rotation rate. For comparison the difference between the model radius of NLTT5306B and the radius for an intermediate gravity brown dwarf of the same mass is 22 per cent. Therefore the signs of intermediate gravity detected cannot be due to this distortion.

The only way to directly measure a brown dwarf radius is by discovering them in eclipsing systems. There are only three known totally eclipsing post-common envelope systems comprising a white dwarf and a brown dwarf, SDSS1411+2009 \citep{littlefair14}, SDSS1205-0242 \citep{parsons17} and WD1032+011 \citep{Casewell20}. SDSS1411+2009 and SDSS1205-0242 have brown dwarf secondaries with similar masses to  NLTT5306B: 0.050$\pm$0.002 M$_{\odot}$ and 0.049$\pm$0.006 M$_{\odot}$ respectively, while WD1032+011 has a much high mass brown dwarf secondary (0.067$\pm$0.006 M$_{\odot}$. These systems have considerably different levels of irradiation however. SDSS1411+2009B (white dwarf T$_{\rm eff}$=13000 K, period of $\sim$ 2 hrs) experiences $\sim$4.5 times the irradiation of NLTT5306B, while the much shorter period SDSS1205-0242B orbiting a white dwarf more than 3 times hotter than NLTT5306A (white dwarf T$_{\rm eff}$=23680$\pm$430 K, period of 71.2 mins) experiences $\sim250$ times the irradiation of NLTT5306B. Both of these systems have masses and radii that are consistent with the BT-Settl models \citep{baraffe15} at ages greater than 2 Gyr, indicating they are not inflated. Both also have much hotter primary stars than NLTT5306A, and as SDSS1411+2009B has a period much longer than that of NLTT5306B ($\sim$ 2 hours), and SDSS1205-0242B has a period much shorter ($\sim$72 min), it is clear that we cannot attribute any inflation in radius of NLTT5306B to heating from the primary star. Indeed, WD1032+011B has a much cooler primary star ($\sim$10000~K), similar to NLTT5306A, experiencing $\sim$1.5 times the irradiation of NLTT5306B, and yet the brown dwarf is inflated \citep{Casewell20}. 
However, there are known brown dwarfs orbiting main sequence stars with similar temperatures to NLTT5306A, for instance HATS-70b \citep{zhou20} and TOI-503b \citep{subjak20}, HATS-70b is inflated, but is a low mass brown dwarf experiencing 80 times the irradiation of NLTT5306B, and TOI-503b is known to be young $\sim$180~Myr, and experiences only one hundredth of the level of irradiation of NLTT5306B so neither are good comparisons here.

\begin{figure*}
\begin{center}
	\includegraphics[scale=0.45]{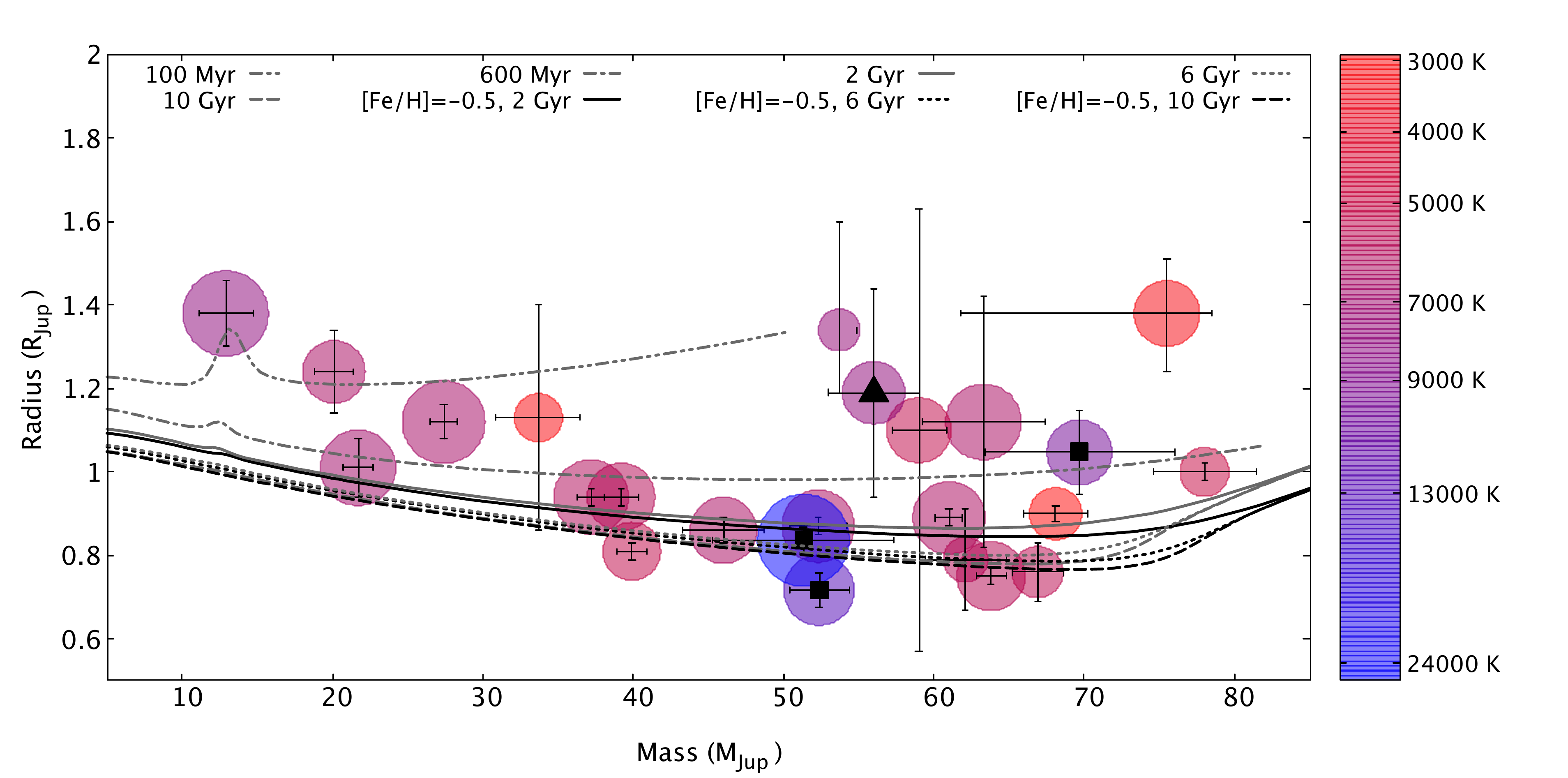}
    \caption{Mass-Radius relation for the 23 known brown dwarfs transiting main sequence stars (\citealt{carmichael} and references therein). The three known brown dwarfs that eclipse white dwarfs \citep{Casewell20, littlefair14, parsons17} are also shown plotted with filled squares. NLTT5306B is shown as a filled triangle.  The effective temperature of the primary star is indicated by the coloured circle, and the size of the coloured circle for each object is proportional to the total irradiation incident on the brown dwarf surface. Also shown for reference are the Sonora-Bobcat evolutionary models of \citet{sonora_web} for 100 Myr, 600 Myr, 2 Gyr, 6 Gyr and 10 Gyr in solar metallicity (grey) and low metallicity ([Fe/H]=-0.5) models for 2 Gyr, 6 Gyr and 10 Gyr (black).}
   \label{fig:radius}
   \end{center}
\end{figure*}

Figure \ref{fig:radius} shows the mass-radius relation for the known 23 brown dwarfs transiting main sequence stars \citep{carmichael20}, as well as the three brown dwarfs that eclipse their white dwarf primaries. NLTT5306B is shown assuming it has a radius of its Roche lobe (in reality, it may of course be smaller than this).  NLTT5306B sits between the 100 Myr and 1 Gyr track.  Perhaps most interestingly, it appears that irrelevant of the amount of irradiation, low mass irradiated brown dwarfs (M$<$35 M$_{\odot}$) \textbf{can} inflate when heated. Higher mass brown dwarfs are however much more difficult to inflate, and the majority are not inflated. In reality, the radius shown for NLTT5306B must be an overestimate as the rate of accretion given in \citet{longstaff18} does not indicate that the brown dwarf is filling its Roche lobe, however, it is clear from our spectrum that the brown dwarf is inflated, just to a lower degree.

To investigate this effect further, we calculated the luminosity of the primary stars in each of the systems in Figure \ref{fig:radius} and the resultant flux at the brown dwarf surface. Using flux means we can compare the irradiation on the brown dwarf surface, taking into account the effective temperature and size of the primary star, as well as the orbital separation. It should be noted however that the effective temperature of the primary star drives the wavelength of the irradiation at the brown dwarf surface, a 10 000~K star will radiate much more in the ultraviolet than a 7 000~K star. There are two systems that receive a similar amount of total irradiation from their primary stars as NLTT5306B and are a similar temperature to NLTT5306A - Kepler-39b (orbits a T$_{\rm eff}$=6350 ~K F7 V: \citealt{bonomo15}) and CoRoT-33b (orbits a 5525~K G9 V: \citealt{cszimadia}).   Kepler-39b has a similar radius to NLTT5306B, but has a much lower mass at $\sim$19 M$_{\rm Jup}$ \citep{bouchy11, bonomo15}.  CoRoT-33b has a similar mass and radius as NLTT5306B as well as a similar amount of irradiation from its host star.  The primary star is metal rich, heavily spotted, and rotates faster than is normal for its G9V spectral type \citep{cszimadia}. However, due to a poorly sampled lightcurve, the radius of the brown dwarf has large errors.   The other higher mass object that appears to have a similar level of inflation to NLTT5306B is CoRoT-15b. This brown dwarf receives 13 times as much flux as NLTT5306B. It should be noted there are other high mass brown dwarfs that receive large amounts of external flux - WASP-30b, WASp-128b and SDSS~J120515.80-024222.6B that are not significantly inflated. CoRoT-15b also orbits an active star.  For comparison, the brown dwarf NLTT41135b receives a much lower level of flux - 0.065 times that of NLTT5306 -  and orbiting an M dwarf, the majority of the irradiation is at a longer wavelength,  and yet with a mass of $\sim$33 M$_{\rm Jup}$ is still inflated indicating that perhaps for low mass objects the level, or wavelength of irradiation is not important to inflate them.

In fact, CoRoT-15b and CoRoT-33b appear to be the only two of these main sequence-brown dwarf systems mentioned to have active primary stars. It is particularly interesting that these three objects are inflated and there is evidence that their host stars have higher than average magnetic fields, this would indicate brown dwarfs may be inflated via a similar mechanism to M dwarfs which also have a large diversity in radii (e.g. \citealt{parsons18}).  Polarimetry measurements of NLTT5306A and WD1032+011A may reveal if they have higher than average magnetic fields (e.g. \citealt{bagnulo}), although neither show any Zeeman splitting in their Balmer lines, suggesting this field is not exceptionally strong and is likely to be less than 1.5-75 MG \citep{bagnulo}.

There has been some discussion regarding Hot Jupiters and whether they may be re-inflated due to the evolution of their host star (e.g. \citealt{komacek, lopez}). For these exoplanets it has been determined that there needs to be a slow deposition of heat into the core over a long period of time, as merely slowing the rate of contraction for the exoplanet is not sufficient to cause the large radii seen for some exoplanets \citep{thorngren, komacek}.  NLTT5306B has an equilibrium temperature of 920~K if albedo is neglected; just on the limit of what is presumed to be possible as exoplanets with T$_{eq}<$1000~K are not found to be inflated. However, brown dwarfs have a significantly higher internal temperature than exoplanets which may have an effect on how the radius of a brown dwarf reacts to heating. The current configuration of the binary has been in place for $\sim$700 Myr, considerably shorter than the 5-10~Gyr reinflation timescales in \citet{komacek} for heat deposited anywhere but the core of the planet. This would suggest if reinflation is the mechanism that inflated the radius of NLTT5306B, then the heat is being deposited right to the centre of the brown dwarf.  This is perhaps consistent with NLTT5306B and WD1032+011B being subject to longer wavelength radiation from the cooler white dwarf than is seen in some of the closer systems which irradiate their brown dwarf companions with considerably more UV photons.  To test this hypothesis we integrated blackbody spectra over the wavelength ranges 0-0.5 $\mu$m, 0.5-1 $\mu$m, 1-2.5 $\mu$m and 2.5-5 $\mu$m for all the the primary stars in Figure \ref{fig:radius}, and then scaled the results to take into account the radius of the primary and orbital separation. No correlation was found between inflation and levels of UV, optical, near-IR or mid-IR irradiation.

Another alternative theory is that a thicker cloud layer may result in a larger radius, or higher metallicity \citep{burrows11}. Both NLTT5306B and WD1032+011B are mid-L dwarfs, and so are expected to be cloudy, but kinematics and cooling ages suggest they are unlikely to be metal enriched, although it is not possible to make a metallicity measurement from such a cool white dwarf as all the heavy elements will have sunk to the core at this age (e.g. \citealt{barstow}).

We conclude the inflation of the higher mass brown dwarfs is most likely to be due to magnetic interactions between the brown dwarf and the primary, as is seen in some systems with an M dwarf secondary,  but we cannot rule out thicker clouds as having an effect. 

\section{Conclusions}
We have shown that NLTT5306B, an irradiated brown dwarf that is weakly interacting with its white dwarf primary star has a low, if any, day-night side brightness temperature difference. However, the brown dwarf is likely to be inflated despite the low temperature ($\sim$7000~K) of its host star. We discuss this inflation in the context of unirradiated brown dwarfs, and also in the context of irradiated brown dwarfs  orbiting main sequence stars. We determine that the irradiation of the brown dwarf CoRoT-33b is extremely similar to that of NLTT5306B, and that both are inflated and orbiting primary stars with higher than average magnetic fields which may be a factor in their inflation.
\section{Data Availability Statement}
The data underlying this article are available in the relevant telescope archives, IRTF and Spitzer, and can be accessed at https://irsa.ipac.caltech.edu/applications/irtf/ and https://sha.ipac.caltech.edu/applications/Spitzer/SHA/ using the object name, co-ordinates or programme number given in the text. The data will be shared on reasonable request to the corresponding author.

\section*{Acknowledgements}
We thank Pierre Bergeron for kindly providing data from \citet{holberg} in the $Spitzer$ bandpasses, and Katelyn Allers for providing her spectral indices code. 
We also thank Richard Alexander for useful discussions regarding circumbinary disks.

SLC acknowledges support from an STFC Ernest Rutherford Fellowship (ST/R003726/1).

These data were taken by a visiting Astronomer at the Infrared Telescope Facility, which is operated by the University of Hawaii under contract NNH14CK55B with the National Aeronautics and Space Administration. This work is also based on observations made with the $Spitzer$ Space Telescope, which is operated by the Jet Propulsion Laboratory, California Institute of Technology under a contract with NASA. This publication makes use of data products from the Wide-field Infrared Survey Explorer, which is a joint project of the University of California, Los Angeles, and the Jet Propulsion Laboratory/California Institute of Technology, funded by the National Aeronautics and Space Administration

We have used the SPLAT routines as part of this work. SPLAT is an experimental, collaborative project of research students in the UCSD Cool Star Lab, aimed at teaching students how to do research by building their own analysis tools. Contributors to SPLAT have included Christian Aganze, Jessica Birky, Daniella Bardalez Gagliuffi, Adam Burgasser (PI), Caleb Choban, Andrew Davis, Ivanna Escala, Joshua Hazlett, Carolina Herrara Hernandez, Elizabeth Moreno Hilario, Aishwarya Iyer, Yuhui Jin, Mike Lopez, Dorsa Majidi, Diego Octavio Talavera Maya, Alex Mendez, Gretel Mercado, Niana Mohammed, Johnny Parra, Maitrayee Sahi, Adrian Suarez, Melisa Tallis, Tomoki Tamiya, Chris Theissen, and Russell van Linge.The SPLAT project is supported by the National Aeronautics and Space Administration under Grant No. NNX15AI75G.




\bibliographystyle{mnras}
\bibliography{bib} 




\bsp	
\label{lastpage}
\end{document}